\documentclass{ptephy}

\usepackage{amssymb}
\usepackage{amsthm}
\usepackage{amsmath}
\usepackage{booktabs}

\numberwithin{equation}{section}

\usepackage{url}

\begin{document}

\title{Energy--momentum tensor from the Yang--Mills gradient flow}

\author{\name{\fname{Hiroshi} \surname{Suzuki}}{\ast}}

\address{Theoretical Research Division, RIKEN Nishina Center, Wako
2-1, Saitama 351-0198, Japan
\email{hsuzuki@riken.jp}}

\begin{abstract}
The product of gauge fields generated by the Yang--Mills gradient flow for
positive flow times does not exhibit the coincidence-point singularity and a
local product is thus independent of the regularization. Such a local product
can furthermore be expanded by renormalized local operators at zero flow time
with finite coefficients that are governed by renormalization group equations.
Using these facts, we derive a formula that relates the small flow-time
behavior of certain gauge-invariant local products and the correctly-normalized
conserved energy--momentum tensor in the Yang--Mills theory. Our formula
provides a possible method to compute the correlation functions of a
well-defined energy--momentum tensor by using lattice regularization and Monte
Carlo simulation.
\end{abstract}

\subjectindex{B01, B31, B32, B38}
\maketitle

\section{Introduction}
\label{sec:1}
Although lattice regularization provides a very powerful non-perturbative
formulation of field theories, it is unfortunately incompatible with
fundamental global symmetries quite often. The most well-known example is
chiral symmetry~\cite{Nielsen:1980rz,Nielsen:1981xu}; supersymmetry is another
infamous example~\cite{Dondi:1976tx}, as is, needless to say, translational
invariance. When a regularization is not invariant under a symmetry, it is not
straightforward to construct the corresponding Noether current that is
conserved and generates the symmetry transformation through Ward--Takahashi
(WT) relations. This makes the measurement of physical quantities related to
the Noether current in a solid basis very difficult. To solve this problem, one
can imagine at least three possible approaches.

The first approach is an ideal one: One finds a lattice formulation that
realizes (a lattice-modified form of) the desired symmetry. If such a
formulation comes to hand, the corresponding Noether current can easily be
obtained by the standard Noether method. The best successful example of this
sort is the lattice chiral symmetry~\cite{Kaplan:1992bt,Neuberger:1997fp,%
Hasenfratz:1997ft,Hasenfratz:1998ri,Neuberger:1998wv,Hasenfratz:1998jp,%
Luscher:1998pqa,Niedermayer:1998bi}, which can be defined with a lattice Dirac
operator that satisfies the Ginsparg--Wilson relation~\cite{Ginsparg:1981bj}.
Although this is certainly an ideal approach, it appears that such an ideal
formulation does not always come to hand, especially for spacetime symmetries
(see, e.g., Ref.~\cite{Kato:2008sp} for a no-go theorem for supersymmetry).

The second approach is to construct the Noether current by tuning coefficients
in the linear combination of operators that can mix with the Noether current
under lattice symmetries.\footnote{Here, we assume that fine tuning of
bare parameters to the target (symmetric) theory is done.} For example, for the
energy--momentum tensor---the Noether current associated with the translational
invariance and rotational and conformal
symmetries~\cite{Callan:1970ze,Coleman:1970je}---one can construct a conserved
lattice energy--momentum tensor by adjusting coefficients in the linear
combination of dimension~$4$
operators~\cite{Caracciolo:1988hc,Caracciolo:1989pt}\footnote{A somewhat
different approach on the basis of the $\mathcal{N}=1$ supersymmetry has been
given in~Refs.~\cite{Suzuki:2013gi,Suzuki:2012wx}.}; the overall normalization
of the energy--momentum tensor has to be fixed in some other way.\footnote{It
might be possible to employ ``current algebra'' for this, as for the axial
current~\cite{Bochicchio:1985xa}.} Although this method is in principle
sufficient when the energy--momentum tensor is in ``isolation'', i.e., when the
energy--momentum tensor is separated from other composite operators, as in the
on-shell matrix elements, it is not obvious a priori whether one can control
the ambiguity of possible higher-dimensional operators that may contribute when
the energy--momentum tensor coincides with other composite operators in
position space. This implies that it is not obvious whether the
energy--momentum tensor constructed in the above method generates
correctly-normalized translations (and rotational and conformal
transformations) on operators through WT relations. (If the energy--momentum
tensor generates correctly-normalized translations, it is
ensured~\cite{Fujikawa:1980rc} (see also Sect.~7.3
of~Ref.~\cite{Fujikawa:2004cx}) that the trace or conformal
anomaly~\cite{Crewther:1972kn,Chanowitz:1972vd} is proportional to the
renormalization group
functions~\cite{Adler:1976zt,Nielsen:1977sy,Collins:1976yq}.)

The third possible approach is to utilize some ultraviolet (UV) finite
quantity. Since such a quantity must be independent of the regularization
adopted (in the limit in which the regulator is removed), there emerges a
possibility that one can relate the lattice regularization and some other
regularization that preserves the desired symmetry. This methodology can be
found e.g.\ in~Ref.~\cite{Luscher:2004fu} (see also
Ref.~\cite{Luscher:2010ik}), where an ultraviolet finite representation of the
topological susceptibility is derived. Although the derivation of the
representation itself relies on a lattice regularization that preserves the
chiral symmetry~\cite{Kaplan:1992bt,Neuberger:1997fp,Hasenfratz:1997ft,%
Hasenfratz:1998ri,Neuberger:1998wv,Hasenfratz:1998jp,Luscher:1998pqa,%
Niedermayer:1998bi}, one can use any regularization (e.g., the Wilson
fermion~\cite{Wilson:1975hf}) to compute the representation because it must be
independent of the regularization.

In the present paper, we consider the above third approach for the
energy--momentum tensor, by taking the pure Yang--Mills theory as an example.
For this, we utilize the so-called Yang--Mills gradient flow (or the Wilson
flow in the context of lattice gauge theory) whose usefulness in lattice gauge
theory has recently been revealed~\cite{Luscher:2010iy,Luscher:2010we,%
Luscher:2011bx,Borsanyi:2012zs,Borsanyi:2012zr,Fodor:2012td,Fodor:2012qh,%
Fritzsch:2013je,Luscher:2013cpa}. A salient feature of the Yang--Mills gradient
flow is its robust UV finiteness~\cite{Luscher:2011bx}. More precisely, any
product of gauge fields generated by the gradient flow for a positive flow
time~$t$ is UV finite under standard renormalization. Such a product, moreover,
does \emph{not\/} exhibit any singularities even if some positions of gauge
fields coincide. The basic mechanism for this UV finiteness is that the flow
equation is a type of the diffusion equation and the evolution operator in the
momentum space~$\sim e^{-tk^2}$ acts as an UV regulator for~$t>0$. This property
of the gradient flow implies that the definition of a local product of gauge
fields for positive flow times is independent of the regularization. In our
present context, there is a hope of relating quantities obtained by the lattice
regularization and the dimensional regularization with which the translational
invariance \emph{is\/} manifest.

As noted in~Ref.~\cite{Luscher:2011bx}, on the other hand, a local product of
gauge fields for a positive flow time can be expanded by renormalized local
operators of the original gauge theory with finite coefficients. Those
coefficients satisfy certain renormalization group equations that, combined
with the dimensional analysis, provide information on the coefficients as a
function of the flow time. Because of the asymptotic freedom, one can then
use the perturbation theory to find the asymptotic behavior of the coefficients
for small flow times.

By using the above properties of the gradient flow, one can obtain a formula
that relates the small flow-time behavior of certain gauge-invariant local
products and the energy--momentum tensor defined by the dimensional
regularization. Since the former can be computed by using the Wilson flow with
lattice regularization~\cite{Luscher:2010iy,Luscher:2010we,Luscher:2011bx,%
Borsanyi:2012zs,Borsanyi:2012zr,Fodor:2012td,Fodor:2012qh,Fritzsch:2013je,%
Luscher:2013cpa} and the latter is conserved and generates correctly-normalized
translations on composite operators, our formula provides a possible method to
compute the correlation functions of a correctly-normalized conserved
energy--momentum tensor by using Monte Carlo simulation.

In the present paper, we follow the notational convention
of~Ref.~\cite{Luscher:2011bx} unless otherwise stated.

\section{Yang--Mills theory and the energy--momentum tensor}
\label{sec:2}
\subsection{The energy--momentum tensor with dimensional regularization}
In the present paper, we consider the $SU(N)$ Yang--Mills theory defined in a
$D$~dimensional Euclidean space. The action is given by
\begin{equation}
   S=\frac{1}{4g_0^2}\int d^Dx\,F_{\mu\nu}^a(x)F_{\mu\nu}^a(x),
\label{eq:(2.1)}
\end{equation}
from the Yang--Mills field strength
\begin{equation}
   F_{\mu\nu}(x)
   =\partial_\mu A_\nu(x)-\partial_\nu A_\mu(x)+[A_\mu(x),A_\nu(x)].
\label{eq:(2.2)}
\end{equation}
We set
\begin{equation}
   D=4-2\epsilon,
\label{eq:(2.3)}
\end{equation}
and then the mass dimension of the bare gauge coupling~$g_0$ is~$\epsilon$.

Assuming that the theory is regularized by the dimensional regularization (for
a very nice exposition, see Ref.~\cite{Collins:1984xc}), one can define the
energy--momentum tensor for the system~\eqref{eq:(2.1)} simply by (see, e.g.,
Ref.~\cite{Freedman:1974gs})
\begin{equation}
   T_{\mu\nu}(x)=\frac{1}{g_0^2}
   \left[F_{\mu\rho}^a(x)F_{\nu\rho}^a(x)
   -\frac{1}{4}\delta_{\mu\nu}F_{\rho\sigma}^a(x)F_{\rho\sigma}^a(x)\right],
\label{eq:(2.4)}
\end{equation}
up to terms attributed to the gauge fixing and the Faddeev--Popov ghost fields,
which are irrelevant in correlation functions of gauge-invariant operators.
Note that the mass dimension of the energy--momentum tensor is~$D$.

The advantage of dimensional regularization is its translational invariance.
Because of this property, the energy--momentum tensor naively constructed from
bare quantities, Eq.~\eqref{eq:(2.4)}, is conserved and generates
correctly-normalized translations through a WT relation,
\begin{equation}
   \int d^Dx\,\left\langle\partial_\mu T_{\mu\nu}(x)\mathcal{O}\right\rangle
   =-\left\langle\partial_\nu\mathcal{O}\right\rangle,
\label{eq:(2.5)}
\end{equation}
where it is understood that the derivative on the right-hand side is acting all
positions in a gauge-invariant operator~$\mathcal{O}$. Used in combination with
dimensional counting and gauge invariance, this WT relation implies that the
energy--momentum tensor~$T_{\mu\nu}(x)$ is
finite~\cite{Joglekar:1975jm,Nielsen:1977sy} and thus, in the minimal
subtraction (MS) scheme,\footnote{Here, we define the renormalized operator by
subtracting its vacuum expectation value. In the perturbation theory using
dimensional regularization, this subtraction is automatic.}
\begin{equation}
   T_{\mu\nu}(x)-\left\langle T_{\mu\nu}(x)\right\rangle
   =\left\{T_{\mu\nu}\right\}_R(x).
\label{eq:(2.6)}
\end{equation}

The finiteness of the energy--momentum tensor~\eqref{eq:(2.4)} provides further
useful information on the renormalization of dimension~$4$ gauge-invariant
operators. The gauge coupling renormalisation with dimensional regularization
is defined by
\begin{equation}
   g_0^2\equiv\mu^{2\epsilon}g^2Z,
\label{eq:(2.7)}
\end{equation}
where $\mu$ is the renormalization scale and $Z$ is the renormalization factor.
In the MS scheme,
\begin{equation}
   Z=1-\frac{1}{\epsilon}
   \left[b_0g^2+\frac{1}{2}b_1g^4+O(g^6)\right]
   +O\left(\frac{1}{\epsilon^2}\right),
\label{eq:(2.8)}
\end{equation}
and
\begin{equation}
   b_0=\frac{11N}{48\pi^2},\qquad
   b_1=\frac{17N^2}{384\pi^4}.
\label{eq:(2.9)}
\end{equation}
From the rotational invariance that the dimensional regularization keeps, we
see that the operator-renormalization possesses the following
structures:\footnote{Here again, we define renormalized operators by
subtracting their vacuum expectation values.}
\begin{align}
   &F_{\mu\rho}^a(x)F_{\nu\rho}^a(x)
   -\left\langle F_{\mu\rho}^a(x)F_{\nu\rho}^a(x)\right\rangle
\notag\\
   &=Z_T\left\{F_{\mu\rho}^aF_{\nu\rho}^a\right\}_R(x)
   +Z_M\delta_{\mu\nu}\left\{F_{\rho\sigma}^aF_{\rho\sigma}^a\right\}_R(x),   
\label{eq:(2.10)}
\end{align}
and
\begin{equation}
   F_{\rho\sigma}^a(x)F_{\rho\sigma}^a(x)
   -\left\langle F_{\rho\sigma}^a(x)F_{\rho\sigma}^a(x)\right\rangle
   =Z_S\left\{F_{\rho\sigma}^aF_{\rho\sigma}^a\right\}_R(x).
\label{eq:(2.11)}
\end{equation}
Substituting the above relations into~Eqs.~\eqref{eq:(2.4)}
and~\eqref{eq:(2.6)}, we have
\begin{align}
   &\left\{T_{\mu\nu}\right\}_R(x)
\notag\\
   &=\frac{1}{g^2}\mu^{-2\epsilon}
   Z^{-1}\biggl[Z_T\left\{F_{\mu\rho}^aF_{\nu\rho}^a\right\}_R(x)
   -\frac{1}{4}(Z_S-4Z_M)
   \delta_{\mu\nu}\left\{F_{\rho\sigma}^aF_{\rho\sigma}^a\right\}_R(x)
   \biggr].
\label{eq:(2.12)}
\end{align}
Since the left-hand side is finite for~$\epsilon\to0$, in the MS scheme in
which only pole terms are subtracted, we infer (by considering the cases,
$\mu\neq\nu$ and~$\mu=\nu$) that
\begin{equation}
   Z_T=Z=1-b_0g^2\frac{1}{\epsilon}+O(g^4)
\label{eq:(2.13)}
\end{equation}
and
\begin{equation}
   Z_S-4Z_M=Z.
\label{eq:(2.14)}
\end{equation}

\subsection{Implications of the trace anomaly}
Another important property of the energy--momentum tensor~\eqref{eq:(2.4)} is
the trace anomaly~\cite{Adler:1976zt,Nielsen:1977sy,Collins:1976yq},
\begin{equation}
   \delta_{\mu\nu}\left\{T_{\mu\nu}\right\}_R(x)
   =-\frac{\beta}{2g^3}
   \left\{F_{\rho\sigma}^aF_{\rho\sigma}^a\right\}_R(x).
\label{eq:(2.15)}
\end{equation}
By Eq.~\eqref{eq:(2.6)}, this relation is equivalent to
\begin{align}
   \delta_{\mu\nu}
   \left[T_{\mu\nu}(x)-\left\langle T_{\mu\nu}(x)\right\rangle\right]
   &=\epsilon\frac{1}{2g_0^2}
   F_{\rho\sigma}^a(x)F_{\rho\sigma}^a(x)
   -\left\langle\epsilon\frac{1}{2g_0^2}
   F_{\rho\sigma}^a(x)F_{\rho\sigma}^a(x)\right\rangle
\notag\\
   &\xrightarrow{\epsilon\to0}
   -\frac{\beta}{2g^3}
   \left\{F_{\rho\sigma}^aF_{\rho\sigma}^a\right\}_R(x).
\label{eq:(2.16)}
\end{align}
In Eqs.~\eqref{eq:(2.15)} and~\eqref{eq:(2.16)}, $\beta$ denotes the
$\beta$~function for $D=4$, defined by
\begin{equation}
   \beta\equiv\left(\mu\frac{\partial}{\partial\mu}\right)_0g
   =-\frac{1}{2}g\left(\mu\frac{\partial}{\partial\mu}\right)_0\ln Z,
\label{eq:(2.17)}
\end{equation}
where the subscript~$0$ implies that the derivative is taken while the bare
quantities are kept fixed. Equations~\eqref{eq:(2.8)} and~\eqref{eq:(2.7)}
yield
\begin{equation}
   \beta=-b_0g^3-b_1g^5+O(g^7).
\label{eq:(2.18)}
\end{equation}
Then, substituting Eqs.~\eqref{eq:(2.12)} into~Eq.~\eqref{eq:(2.15)} and using
Eqs.~\eqref{eq:(2.13)} and~\eqref{eq:(2.14)}, we observe that
\begin{equation}
   \delta_{\rho\lambda}
   \left\{F_{\rho\sigma}^aF_{\lambda\sigma}^a\right\}_R(x)
   =\left(1-\frac{\beta}{2g}\right)
   \left\{F_{\rho\sigma}^aF_{\rho\sigma}^a\right\}_R(x),
\label{eq:(2.19)}
\end{equation}
i.e., the contraction with the metric and the minimal subtraction, the
subtraction of $1/\epsilon$~poles, do not commute; this is a peculiar but
legitimate property of the dimensional regularization~\cite{Collins:1984xc}.

Also, substituting Eqs.~\eqref{eq:(2.7)} and~\eqref{eq:(2.11)}
into~Eq.~\eqref{eq:(2.16)}, we see
\begin{equation}
   \epsilon\frac{Z_S}{Z}
   \xrightarrow{\epsilon\to0}
   -\frac{\beta}{g}.
\label{eq:(2.20)}
\end{equation}
In the MS scheme in which only pole terms are subtracted, this implies
\begin{equation}
   Z_S=\left(1-\frac{\beta}{g}\frac{1}{\epsilon}\right)Z
   =1+O(g^4),
\label{eq:(2.21)}
\end{equation}
and Eq.~\eqref{eq:(2.14)} then shows
\begin{equation}
   Z_M=-\frac{\beta}{4g}Z\frac{1}{\epsilon}
   =\frac{b_0}{4}g^2\frac{1}{\epsilon}+O(g^4).
\label{eq:(2.22)}
\end{equation}
We thus observe that all the renormalization constants
in~Eqs.~\eqref{eq:(2.10)} and~\eqref{eq:(2.11)}, $Z_T$, $Z_M$ and~$Z_S$, in the
MS scheme can eventually be expressed by the gauge coupling renormalization
constant~$Z$ in~Eq.~\eqref{eq:(2.7)}.

\section{Yang--Mills gradient flow and the small flow-time expansion}
\label{sec:3}
The Yang--Mills gradient flow defines a $D+1$~dimensional gauge
potential~$B(t,x)$ along a fictitious time~$t$, according to the flow equation
\begin{equation}
   \partial_tB_\mu(t,x)=D_\nu G_{\nu\mu}(t,x)
   +\alpha_0D_\mu\partial_\nu B_\nu(t,x),
\label{eq:(3.1)}
\end{equation}
where the $D+1$~dimensional field strength and the covariant derivative are
defined by
\begin{equation}
   G_{\mu\nu}(t,x)
   =\partial_\mu B_\nu(t,x)-\partial_\nu B_\mu(t,x)
   +[B_\mu(t,x),B_\nu(t,x)]
\label{eq:(3.2)}
\end{equation}
and
\begin{equation}
   D_\mu=\partial_\mu+[B_\mu,\cdot],
\label{eq:(3.3)}
\end{equation}
respectively. The initial condition for the flow is given by the
$D$~dimensional gauge potential in the previous section:
\begin{equation}
   B_\mu(t=0,x)=A_\mu(x).
\label{eq:(3.4)}
\end{equation}
In Eq.~\eqref{eq:(3.1)}, the last term is introduced to suppress the evolution
of the field along the direction of gauge degrees of freedom. Although this
term breaks the gauge symmetry, it does not affect the evolution of any
gauge-invariant operators~\cite{Luscher:2010iy}. Note that the mass dimension
of the flow time~$t$ is~$-2$.

Now, from the field strength extended to the $D+1$~dimension~\eqref{eq:(3.2)},
we define a $D+1$~dimensional analogue of the energy--momentum tensor by
\begin{equation}
   U_{\mu\nu}(t,x)\equiv
   G_{\mu\rho}^a(t,x)G_{\nu\rho}^a(t,x)
   -\frac{1}{4}\delta_{\mu\nu}G_{\rho\sigma}^a(t,x)G_{\rho\sigma}^a(t,x).
\label{eq:(3.5)}
\end{equation}
Although this is similar in form to the original energy--momentum
tensor~\eqref{eq:(2.4)}, it is not obvious a priori how this $D+1$~dimensional
object and Eq.~\eqref{eq:(2.4)} are related (or not). To find the relationship
between them is the principal task of the present paper. We also use the
density operator studied in~Ref.~\cite{Luscher:2010iy}:
\begin{equation}
   E(t,x)\equiv\frac{1}{4}G_{\mu\nu}^a(t,x)G_{\mu\nu}^a(t,x).
\label{eq:(3.6)}
\end{equation}

Now, as shown in~Ref.~\cite{Luscher:2011bx}, for~$t>0$, any correlation
function of~$B_\mu(t,x)$ is UV finite after standard renormalization in the
$D$~dimensional Yang--Mills theory. This property holds even for any local
products of~$B_\mu(t,x)$ such as~Eqs.~\eqref{eq:(3.5)} and~\eqref{eq:(3.6)}.
Also, for small flow times, a local product of~$B_\mu(t,x)$ can be regarded as
a local field in the $D$~dimensional sense because the flow
equation~\eqref{eq:(3.1)} is basically the diffusion equation along the
time~$t$ and the diffusion length in~$x$ is~$\sqrt{8t}$. These properties allow
us to express, as explained in~Sect.~8 of~Ref.~\cite{Luscher:2011bx},
$U_{\mu\nu}(t,x)$ and~$E(t,x)$ as an asymptotic series of $D$~dimensional
renormalized local operators with finite coefficients. Considering the gauge
invariance and the index structure, for $D=4$, we can write
\begin{equation}
   U_{\mu\nu}(t,x)
   =c_T(t)\left\{T_{\mu\nu}\right\}_R(x)
   +c_S(t)\delta_{\mu\nu}
   \left\{\frac{1}{4}F_{\rho\sigma}^aF_{\rho\sigma}^a\right\}_R(x)
   +O(t),
\label{eq:(3.7)}
\end{equation}
where abbreviated terms are the contributions of operators with a mass
dimension higher than or equal to~$6$. For Eq.~\eqref{eq:(3.6)}, we similarly
have
\begin{equation}
   E(t,x)
   =\left\langle E(t,x)\right\rangle
   +c_E(t)\left\{\frac{1}{4}F_{\rho\sigma}^aF_{\rho\sigma}^a\right\}_R(x)
   +O(t).
\label{eq:(3.8)}
\end{equation}
We note that, when the renormalized gauge coupling is fixed,
$U_{\mu\nu}(t,x)$~\eqref{eq:(3.5)} is traceless for~$D=4$,
\begin{equation}
   \delta_{\mu\nu}U_{\mu\nu}(t,x)
   =2\epsilon E(t,x)\xrightarrow{\epsilon\to0}0,
\label{eq:(3.9)}
\end{equation}
because $E(t,x)$~\eqref{eq:(3.6)} is finite~\cite{Luscher:2010iy} and does not
produce a $1/\epsilon$ singularity (this explains why there is no $c$~number
expectation value term in~Eq.~\eqref{eq:(3.7)}). Thus, considering the trace
part of~Eq.~\eqref{eq:(3.7)}, we see that the coefficients $c_T(t)$
and~$c_S(t)$ are not independent and are related by, for~$D=4$,
\begin{equation}
   c_S(t)=\frac{\beta}{2g^3}c_T(t),
\label{eq:(3.10)}
\end{equation}
because of the trace anomaly~\eqref{eq:(2.15)}.

By eliminating the renormalized action density from Eqs.~\eqref{eq:(3.7)}
and~\eqref{eq:(3.8)}, we have
\begin{equation}
   \left\{T_{\mu\nu}\right\}_R(x)
   =\frac{1}{c_T(t)}U_{\mu\nu}(t,x)
   -\frac{c_S(t)}{c_T(t)c_E(t)}\delta_{\mu\nu}
   \left[E(t,x)
   -\left\langle E(t,x)\right\rangle
   \right]
   +O(t).
\label{eq:(3.11)}
\end{equation}
This expression relates the energy--momentum tensor~\eqref{eq:(2.6)} and the
short flow-time behavior of gauge-invariant local products defined by the
gradient flow. Thus, once the coefficients are known, one can extract the
energy--momentum tensor from the $t\to0$ behavior of the combination on the
right-hand side.

\section{Renormalization group equation and the asymptotic formula}
\label{sec:4}
\subsection{Renormalization group equation for the coefficients}
\label{sec:4.1}
We now operate
\begin{equation}
   \left(\mu\frac{\partial}{\partial\mu}\right)_0,
\label{eq:(4.1)}
\end{equation}
on both sides of Eq.~\eqref{eq:(3.7)}. Since the left-hand side
of~Eq.~\eqref{eq:(3.7)}, i.e., Eq.~\eqref{eq:(3.5)}, is entirely expressed by
bare quantities through the flow equation~\eqref{eq:(3.1)} and the initial
condition~\eqref{eq:(3.4)}, the action of~\eqref{eq:(4.1)} on the left-hand
side identically vanishes. On the right-hand side, this vanishing must hold in
each power of~$t$. Thus we infer that
\begin{align}
   &\left(\mu\frac{\partial}{\partial\mu}\right)_0c_T(t)
   \left\{T_{\mu\nu}\right\}_R(x)=0,
\label{eq:(4.2)}
\\
   &\left(\mu\frac{\partial}{\partial\mu}\right)_0c_S(t)
   \left\{\frac{1}{4}F_{\rho\sigma}^aF_{\rho\sigma}^a\right\}_R(x)=0.
\label{eq:(4.3)}
\end{align}

For the first relation~\eqref{eq:(4.2)}, we recall that the energy--momentum
tensor is not renormalized as~Eq.~\eqref{eq:(2.6)}. Then, by expressing the
operation~\eqref{eq:(4.1)} in terms of renormalized quantities, we have
\begin{equation}
   \left(\mu\frac{\partial}{\partial\mu}
   +\beta\frac{\partial}{\partial g}\right)c_T(t)=0.
\label{eq:(4.4)}
\end{equation}

For Eq.~\eqref{eq:(4.3)}, on the other hand, from~Eq.~\eqref{eq:(2.11)},
\begin{equation}
   \left(\mu\frac{\partial}{\partial\mu}
   +\beta\frac{\partial}{\partial g}
   +\gamma_S\right)c_S(t)=0,
\label{eq:(4.5)}
\end{equation}
where
\begin{equation}
   \gamma_S
   \equiv-\left(\mu\frac{\partial}{\partial\mu}\right)_0\ln Z_S.
\label{eq:(4.6)}
\end{equation}
Equations~\eqref{eq:(2.21)}, \eqref{eq:(2.7)}, and~\eqref{eq:(2.17)} yield 
\begin{equation}
   \gamma_S=-g^3\frac{d}{dg}\left(\frac{\beta}{g^3}\right)
   =2b_1g^4+O(g^6).
\label{eq:(4.7)}
\end{equation}

Similarly, for~Eq.~\eqref{eq:(3.8)}, we have
\begin{align}
   &\left(\mu\frac{\partial}{\partial\mu}
   +\beta\frac{\partial}{\partial g}\right)
   \left\langle E(t,x)\right\rangle=0,
\label{eq:(4.8)}
\\
   &\left(\mu\frac{\partial}{\partial\mu}
   +\beta\frac{\partial}{\partial g}
   +\gamma_S\right)c_E(t)=0,
\label{eq:(4.9)}
\end{align}
and thus
\begin{equation}
   \left(\mu\frac{\partial}{\partial\mu}
   +\beta\frac{\partial}{\partial g}\right)\frac{c_S(t)}{c_E(t)}=0.
\label{eq:(4.10)}
\end{equation}

By the standard argument and from the fact that dimensionless quantities can
depend on the renormalization scale~$\mu$ only through the dimensionless
combination~$\sqrt{8t}\mu$, the above renormalization group equations imply
that
\begin{align}
   &c_T(t)(g;\mu)=c_T(t_0)(\Bar{g}(-\xi);\mu_0),
\label{eq:(4.11)}
\\
   &c_S(t)(g;\mu)
   =\exp\left[\int_0^{-\xi}d\xi'\,\gamma_S\left(\Bar{g}(\xi')\right)\right]
   c_S(t_0)(\Bar{g}(-\xi);\mu_0),
\label{eq:(4.12)}
\\
   &t^2\left\langle E(t,x)\right\rangle(g;\mu)
   =t_0^2\left\langle E(t_0,x)\right\rangle(\Bar{g}(-\xi);\mu_0),
\label{eq:(4.13)}
\\
   &\frac{c_S(t)}{c_E(t)}(g;\mu)=\frac{c_S(t_0)}{c_E(t_0)}(\Bar{g}(-\xi);\mu_0),
\label{eq:(4.14)}
\end{align}
where the dependence on the renormalized gauge coupling and on the
renormalization scale has been explicitly written. In these expressions, the
running coupling~$\Bar{g}(\xi)$ is defined by
\begin{equation}
   \frac{d\Bar{g}(\xi)}{d\xi}=\beta\left(\Bar{g}(\xi)\right),\qquad
   \Bar{g}(0)=g,
\label{eq:(4.15)}
\end{equation}
and we introduce a variable
\begin{equation}
   \xi\equiv\ln\frac{\sqrt{8t}\mu}{\sqrt{8t_0}\mu_0}.
\label{eq:(4.16)}
\end{equation}

In the one-loop order, the running couping~\eqref{eq:(4.15)} is given by
\begin{equation}
   \Bar{g}(-\xi)^2=\frac{1}{2b_0}\frac{1}{-\xi+1/(2b_0g^2)}
   =\frac{1}{2b_0}\frac{1}{-\ln(\sqrt{8t}\Lambda)+\ln(\sqrt{8t_0}\mu_0)},
\label{eq:(4.17)}
\end{equation}
where $\Lambda$ is the $\Lambda$~parameter in the one-loop level,
\begin{equation}
   \Lambda=\mu e^{-1/(2b_0g^2)},
\label{eq:(4.18)}
\end{equation}
and the integral appearing in~Eqs.~\eqref{eq:(4.12)} is
\begin{equation}
   \int_0^{-\xi}d\xi'\,
   \gamma_S\left(\Bar{g}(\xi')\right)
   =\frac{b_1}{b_0}
   \left[
   g^2-\Bar{g}(-\xi)^2
   \right].
\label{eq:(4.19)}
\end{equation}
In the small flow-time limit $t\to0$, $-\xi\to+\infty$ and the running
coupling~$\Bar{g}(-\xi)$~\eqref{eq:(4.17)} becomes very small thanks to the
asymptotic freedom. Thus, the right-hand sides
of~Eqs.~\eqref{eq:(4.11)}--\eqref{eq:(4.14)} allow us to compute the small
flow-time behavior of the coefficients by using the perturbation theory.

\subsection{Lowest-order approximation and the asymptotic formula}
By substituting the solution of the flow equation~\eqref{eq:(3.1)} (see
Ref.~\cite{Luscher:2011bx}) in the tree-level approximation
to~Eq.~\eqref{eq:(3.7)}, we have
\begin{equation}
   c_T(t)=g_0^2,
\label{eq:(4.20)}
\end{equation}
simply because our energy--momentum tensor~\eqref{eq:(2.4)} is proportional
to~$1/g_0^2$. If we apply the right-hand side of~Eq.~\eqref{eq:(4.11)} to this
expression by substituting Eq.~\eqref{eq:(4.17)}, however, it depends
on~$\sqrt{8t_0}\mu_0$ while the left-hand side of~Eq.~\eqref{eq:(4.11)} does
not. This shows that $c_T(t)$ should depend on~$g^2$ and~$\sqrt{8t}\mu$ through
a particular combination as (for $D=4$)
\begin{equation}
   c_T(t)=
   g^2\left\{1+2b_0g^2\left[\ln(\sqrt{8t}\mu)+c_1\right]+O(g^4)\right\},
\label{eq:(4.21)}
\end{equation}
where $c_1$ is a constant. Similarly, since the lowest-order approximation
in~Eqs.~\eqref{eq:(3.7)} and~\eqref{eq:(3.8)} yields
\begin{equation}
   c_E(t)=1,\qquad
   c_S(t)=-\frac{b_0}{2}g_0^2\mu^{-2\epsilon},
\label{eq:(4.22)}
\end{equation}
where the latter follows from~Eq.~\eqref{eq:(3.10)}, from~Eq.~\eqref{eq:(4.14)}
we have
\begin{equation}
   \frac{c_S(t)}{c_E(t)}
   =-\frac{b_0}{2}
   g^2\left\{1+2b_0g^2\left[\ln(\sqrt{8t}\mu)+c_2\right]+O(g^4)\right\},
\label{eq:(4.23)}
\end{equation}
where $c_2$ is another constant.\footnote{Using Eq.~\eqref{eq:(4.21)}
in~Eq.~\eqref{eq:(3.10)}, we have
\begin{equation}
   c_S(t)=-\frac{b_0}{2}
   g^2\left\{
   1+2b_0g^2
   \left[
   \ln(\sqrt{8t}\mu)+c_1+\frac{b_1}{2b_0^2}\right]+O(g^4)
   \right\},
\label{eq:(4.24)}
\end{equation}
and then using Eq.~\eqref{eq:(4.23)},
\begin{equation}
   c_E(t)=1+2b_0g^2
   \left(c_1-c_2+\frac{b_1}{2b_0^2}\right)+O(g^4).
\label{eq:(4.25)}
\end{equation}
}

Applying Eqs.~\eqref{eq:(4.11)} and~\eqref{eq:(4.14)} to above expressions
and using~Eq.~\eqref{eq:(4.17)}, we finally have the asymptotic behaviors of
the coefficients in~Eq.~\eqref{eq:(3.11)},
\begin{equation}
   \frac{1}{c_T(t)}
   \stackrel{t\to0+}{\sim}
   -2b_0\left[\ln(\sqrt{8t}\Lambda)+c_1\right]
\label{eq:(4.26)}
\end{equation}
and
\begin{equation} 
   \frac{c_S(t)}{c_E(t)}
   \stackrel{t\to0+}{\sim}
   -\frac{b_0}{2}
   \frac{1}{-2b_0\left[\ln(\sqrt{8t}\Lambda)+c_2\right]},
\label{eq:(4.27)}
\end{equation}
and hence
\begin{equation} 
   \frac{c_S(t)}{c_T(t)c_E(t)}
   \stackrel{t\to0+}{\sim}
   -\frac{b_0}{2}
   \left[1-\frac{c_1-c_2}{-\ln(\sqrt{8t}\Lambda)}\right].
\label{eq:(4.28)}
\end{equation}
That is,
\begin{align}
   \left\{T_{\mu\nu}\right\}_R(x)
   &\stackrel{t\to0+}{\sim}
   \biggl\{
   -2b_0\left[\ln(\sqrt{8t}\Lambda)+c_1\right]U_{\mu\nu}(t,x)
\notag\\
   &\qquad\qquad{}
   +\frac{b_0}{2}
   \left[1-\frac{c_1-c_2}{-\ln(\sqrt{8t}\Lambda)}\right]
   \delta_{\mu\nu}
   \left[
   E(t,x)-\left\langle E(t,x)\right\rangle\right]
   \biggr\}.
\label{eq:(4.29)}
\end{align}
This is the relation that we were seeking: One can obtain the
correctly-normalized conserved energy--momentum tensor from the small flow-time
behavior of gauge-invariant products given by the Yang--Mills gradient flow. It
is interesting to note that the leading $t\to0$~behavior is completely
independent of the detailed definition of the gradient flow; the structure and
coefficients follow solely from the finiteness of the local products and the
renormalizability of the Yang--Mills theory. The sub-leading corrections in the
asymptotic form, i.e., the coefficients $c_1$ and~$c_2$, depend on the detailed
definition of the gradient flow; in the Appendix, we compute the constants
$c_1$ and~$c_2$ and we have
\begin{align}
   c_1&=\ln\sqrt{\pi}+\frac{7}{22}
   \simeq0.890547,
\label{eq:(4.30)}
\\
   c_2&=\ln\sqrt{\pi}-\frac{7}{44}
   +\frac{b_1}{2b_0^2}
   \simeq0.834762.
\label{eq:(4.31)}
\end{align}

Finally, a possible method to determine the factor $\ln(\sqrt{8t}\Lambda)$
in~Eq.~\eqref{eq:(4.29)}, i.e., the flow time~$t$ in the unit of the one-loop
$\Lambda$ parameter~\eqref{eq:(4.18)}, for small flow times is to use the
expectation value of the density operator, Eq.~\eqref{eq:(3.6)}. For this
quantity, by applying Eqs.~\eqref{eq:(4.13)} and~\eqref{eq:(4.17)} to the
result of the one-loop calculation, Eqs.~(2.28) and~(2.29)
of~Ref.~\cite{Luscher:2010iy} (specialized to the pure Yang--Mills theory), we
have the asymptotic form,
\begin{equation}
   t^2\left\langle E(t,x)\right\rangle
   \stackrel{t\to0+}{\sim}
   \frac{3(N^2-1)}{128\pi^2}
   \frac{1}{-2b_0\left[\ln(\sqrt{8t}\Lambda)+c\right]},
\label{eq:(4.32)}
\end{equation}
where
\begin{equation}
   c\equiv\ln(2\sqrt{\pi})+\frac{26}{33}-\frac{9}{22}\ln3
   \simeq1.60396.
\label{eq:(4.33)}
\end{equation}
One may use this asymptotic representation for $\ln(\sqrt{8t}\Lambda)$
in~Eq.~\eqref{eq:(4.29)}.\footnote{In practice, one will use
Eq.~\eqref{eq:(4.29)} to compute $t^2\{T_{\mu\nu}\}_R(x)$
from~$t^2U_{\mu\nu}(t,x)$ and~$t^2E(t,x)$. Then, from the value
of~$\sqrt{8t}\Lambda$, one can deduce $\{T_{\mu\nu}\}_R(x)/\Lambda^4$.}

\section{Conclusion}
In the present paper, we have derived a formula that relates the short
flow-time behavior of some gauge-invariant local products generated by the
Yang--Mills gradient flow and the correctly-normalized conserved
energy--momentum tensor in the Yang--Mills theory. Our main result is
Eq.~\eqref{eq:(4.29)}. The right-hand side of~Eq.~\eqref{eq:(4.29)} can be
computed by the Wilson flow in lattice gauge theory with appropriate
discretizations of operators, Eqs.~\eqref{eq:(3.5)} and~\eqref{eq:(3.6)} (see,
e.g., Refs.~\cite{Luscher:2010iy,Borsanyi:2012zs}). Here, the continuum
limit~$a\to0$ must be taken first and then the $t\to0$ limit is taken
afterwards; otherwise our basic reasoning does not hold.

Although the formula~\eqref{eq:(4.29)} should be mathematically correct, the
practical usefulness of~Eq.~\eqref{eq:(4.29)} is a separate issue and has to be
carefully examined numerically.\footnote{We hope to return to this problem in
the near future.} Since the lattice spacing~$a$ must be sufficiently smaller
than the square-root of the flow time~$\sqrt{8t}$ for our reasoning to work,
the reliable application of~Eq.~\eqref{eq:(4.29)} will require rather small
lattice spacings. One also worries about contamination by higher-dimensional
operators (i.e., the $O(t)$ terms in~Eqs.~\eqref{eq:(3.7)}
and~\eqref{eq:(3.8)}) and the finite-size effect which we have not taken into
account in the present paper. If our strategy turns to be practically feasible,
it provides a completely new method to compute correlation functions containing
a well-defined energy--momentum tensor. It is clear that the present approach
to the energy--momentum tensor on the lattice is not limited to the pure
Yang--Mills theory although the treatment might be slightly more complicated
with the presence of other fields. The application will then include the
determination of the shear and bulk viscosities (see, e.g.,
Refs.~\cite{Meyer:2007ic,Meyer:2007dy}), the measurement of
thermodynamical quantities (see Ref.~\cite{Giusti:2012yj} and references cited
therein), the mass and the decay constant of the pseudo Nambu--Goldstone boson
associated with the (approximate) dilatation invariance (see
Ref.~\cite{Appelquist:2010gy} and references cited therein), and so on.

It is also clear that our basic idea, that operators defined with lattice
regularization and in the continuum theory can be related through the gradient
flow is not limited to the energy--momentum tensor. For example, it might be
possible to construct an ideal chiral current or an ideal supercurrent on the
lattice, from the small flow-time limit of local products. It would be
interesting to pursue this idea.

\section*{Acknowledgements}
The possibility that the Yang--Mills gradient flow (or the Wilson flow) can be
useful for defining the energy--momentum tensor in lattice gauge theory was
originally suggested to me by Etsuko Itou. I would like to thank her for
enlightening discussions. I would also like to thank Martin L\"uscher for a
clarifying remark on the precise meaning of~Eq.~\eqref{eq:(3.8)}.
I am grateful to Hiroki Makino for his help in finding errors in the one-loop
calculation in previous versions of the preset paper.
This work is supported in part by a Grant-in-Aid for Scientific
Research~23540330.

\appendix

\section{One-loop calculation of coefficient functions}
\label{sec:5}
For calculational convenience, we define the coefficient functions $F(t)$
and~$G(t)$ by
\begin{align}
   &G_{\mu\rho}^a(t,x)G_{\nu\rho}^a(t,x)
   -\left\langle G_{\mu\rho}^a(t,x)G_{\nu\rho}^a(t,x)\right\rangle
\notag\\
   &\qquad{}
   =F(t)\left\{F_{\mu\rho}^aF_{\nu\rho}^a\right\}_R(x)
   +G(t)\delta_{\mu\nu}\left\{F_{\rho\sigma}^aF_{\rho\sigma}^a\right\}_R(x)
   +O(t).
\label{eq:(5.1)}
\end{align}
Equation~\eqref{eq:(3.5)} then becomes (for $D=4$),
\begin{align}
   U_{\mu\nu}(t,x)
   &=F(t)\left[
   \left\{F_{\mu\rho}^aF_{\nu\rho}^a\right\}_R(x)
   -\frac{1}{4}\delta_{\mu\nu}
   \delta_{\rho\lambda}\left\{F_{\rho\sigma}^aF_{\lambda\sigma}^a\right\}_R(x)
   \right]
   +O(t)
\notag\\
   &=F(t)\left[
   \left\{F_{\mu\rho}^aF_{\nu\rho}^a\right\}_R(x)
   -\frac{1}{4}\delta_{\mu\nu}
   \left(1-\frac{\beta}{2g}\right)
   \left\{F_{\rho\sigma}^aF_{\rho\sigma}^a\right\}_R(x)
   \right]
   +O(t),
\label{eq:(5.2)}
\end{align}
where we have used Eq.~\eqref{eq:(2.19)}. Rewriting this in favor of the
energy--momentum tensor~\eqref{eq:(2.12)} with Eqs.~\eqref{eq:(2.13)}
and~\eqref{eq:(2.14)}, we have
\begin{align}
   U_{\mu\nu}(t,x)
   &=F(t)\left[
   \left\{F_{\mu\rho}^aF_{\nu\rho}^a\right\}_R(x)
   -\frac{1}{4}\delta_{\mu\nu}
   \delta_{\rho\lambda}\left\{F_{\rho\sigma}^aF_{\lambda\sigma}^a\right\}_R(x)
   \right]
   +O(t)
\notag\\
   &=F(t)\left[
   g^2\left\{T_{\mu\nu}\right\}_R(x)
   +\frac{\beta}{8g}\delta_{\mu\nu}
   \left\{F_{\rho\sigma}^aF_{\rho\sigma}^a\right\}_R(x)
   \right]
   +O(t).
\label{eq:(5.3)}
\end{align}
Comparison with~Eq.~\eqref{eq:(3.7)} then shows
\begin{equation}
   c_T(t)=g^2F(t),\qquad c_S(t)=\frac{\beta}{2g}F(t).
\label{eq:(5.4)}
\end{equation}

Similarly, for~Eq.~\eqref{eq:(3.8)},
\begin{align}
   E(t,x)
   &=\left\langle E(t,x)\right\rangle
   +\frac{1}{4}F(t)\delta_{\rho\lambda}
   \left\{F_{\rho\sigma}^aF_{\lambda\sigma}^a\right\}_R(x)
   +G(t)\left\{F_{\rho\sigma}^aF_{\rho\sigma}^a\right\}_R(x)
   +O(t)
\notag\\
   &=\left[
   \left(1-\frac{\beta}{2g}\right)F(t)+4G(t)\right]
   \left\{\frac{1}{4}F_{\rho\sigma}^aF_{\rho\sigma}^a\right\}_R(x)
   +O(t),
\label{eq:(5.5)}
\end{align}
and therefore
\begin{equation}
   c_E(t)=\left(1-\frac{\beta}{2g}\right)F(t)+4G(t).
\label{eq:(5.6)}
\end{equation}
This implies, for the ratio~\eqref{eq:(4.23)},
\begin{align}
   \frac{c_S(t)}{c_E(t)}
   &=\frac{\beta}{2g}\frac{1}{1-\frac{\beta}{2g}+4G(t)/F(t)}
\notag\\
   &=-\frac{b_0}{2}g^2
   \left\{1+2b_0g^2\left[-\frac{1}{4}+\frac{b_1}{2b_0^2}\right]
   -4G(t)+O(g^4)\right\}.
\label{eq:(5.7)}
\end{align}

\begin{figure}
\centering
\includegraphics[width=3cm,clip]{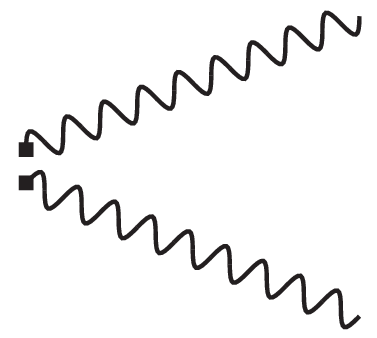}
\caption{}
\label{fig:1}
\end{figure}

\begin{figure}
\centering
\includegraphics[width=3cm,clip]{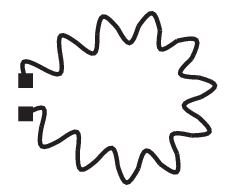}
\caption{}
\label{fig:2}
\end{figure}

\begin{figure}
\begin{minipage}{0.3\hsize}
\begin{center}
\includegraphics[width=3cm,clip]{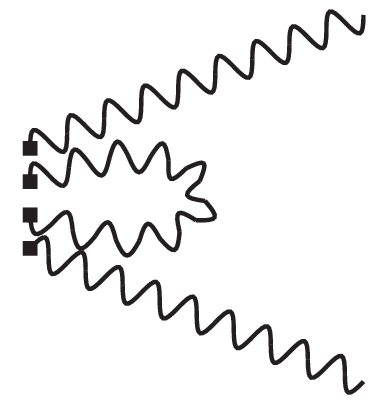}
\caption{}
\label{fig:3}
\end{center}
\end{minipage}
\begin{minipage}{0.3\hsize}
\begin{center}
\includegraphics[width=3cm,clip]{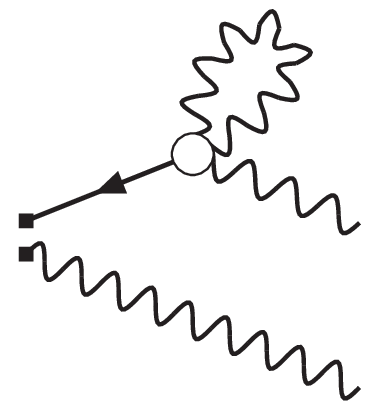}
\caption{}
\label{fig:4}
\end{center}
\end{minipage}
\begin{minipage}{0.3\hsize}
\begin{center}
\includegraphics[width=3cm,clip]{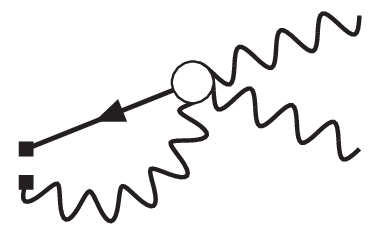}
\caption{}
\label{fig:5}
\end{center}
\end{minipage}
\end{figure}

\begin{figure}
\begin{minipage}{0.3\hsize}
\begin{center}
\includegraphics[width=4cm,clip]{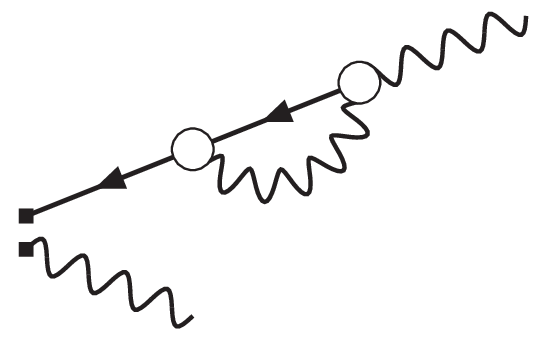}
\caption{}
\label{fig:6}
\end{center}
\end{minipage}
\begin{minipage}{0.3\hsize}
\begin{center}
\includegraphics[width=4cm,clip]{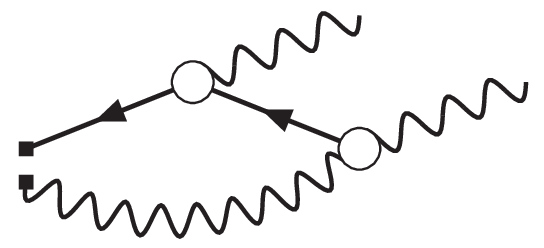}
\caption{}
\label{fig:7}
\end{center}
\end{minipage}
\begin{minipage}{0.3\hsize}
\begin{center}
\includegraphics[width=3cm,clip]{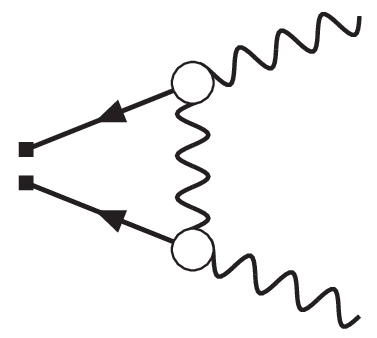}
\caption{}
\label{fig:8}
\end{center}
\end{minipage}
\end{figure}

\begin{figure}
\begin{minipage}{0.3\hsize}
\begin{center}
\includegraphics[width=3cm,clip]{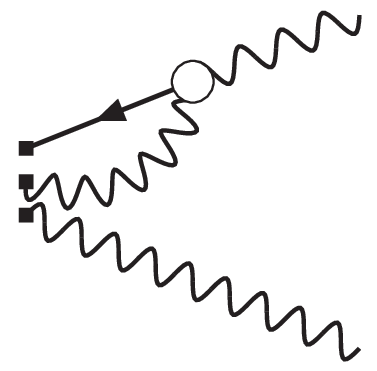}
\caption{}
\label{fig:9}
\end{center}
\end{minipage}
\begin{minipage}{0.3\hsize}
\begin{center}
\includegraphics[width=3cm,clip]{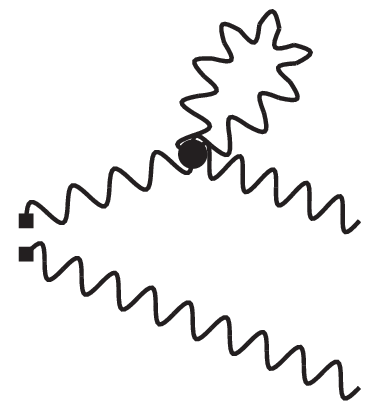}
\caption{}
\label{fig:10}
\end{center}
\end{minipage}
\begin{minipage}{0.3\hsize}
\begin{center}
\includegraphics[width=3cm,clip]{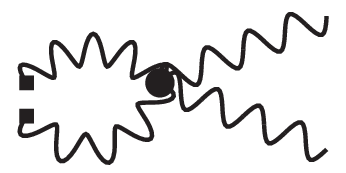}
\caption{}
\label{fig:11}
\end{center}
\end{minipage}
\end{figure}

\begin{figure}
\begin{minipage}{0.3\hsize}
\begin{center}
\includegraphics[width=4cm,clip]{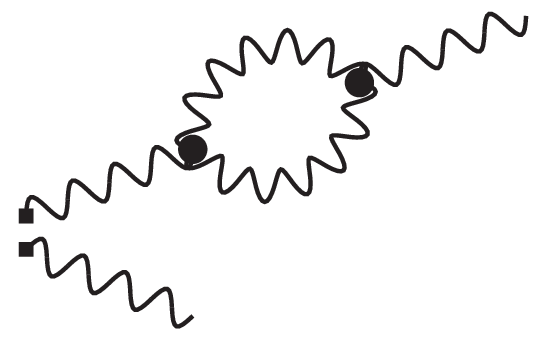}
\caption{}
\label{fig:12}
\end{center}
\end{minipage}
\begin{minipage}{0.3\hsize}
\begin{center}
\includegraphics[width=3cm,clip]{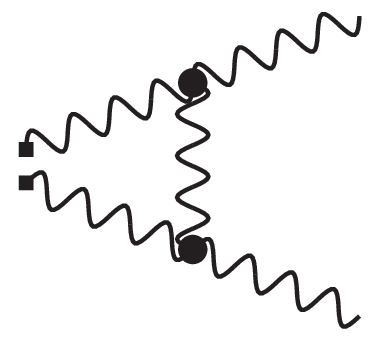}
\caption{}
\label{fig:13}
\end{center}
\end{minipage}
\begin{minipage}{0.3\hsize}
\begin{center}
\includegraphics[width=3cm,clip]{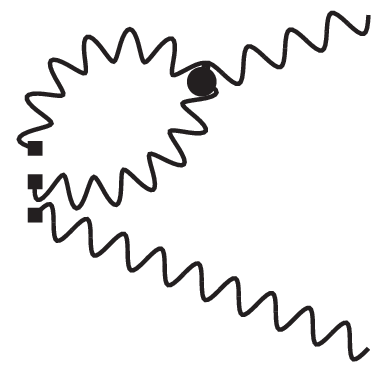}
\caption{}
\label{fig:14}
\end{center}
\end{minipage}
\end{figure}

\begin{figure}
\begin{minipage}{0.3\hsize}
\begin{center}
\includegraphics[width=4cm,clip]{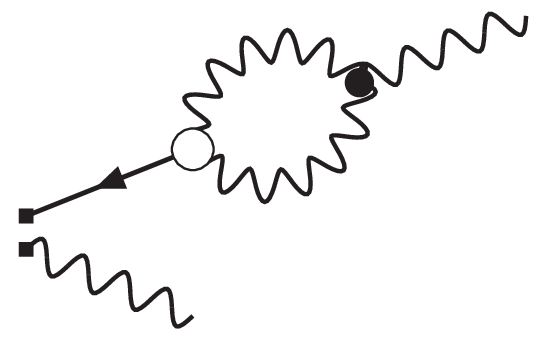}
\caption{}
\label{fig:15}
\end{center}
\end{minipage}
\begin{minipage}{0.3\hsize}
\begin{center}
\includegraphics[width=3cm,clip]{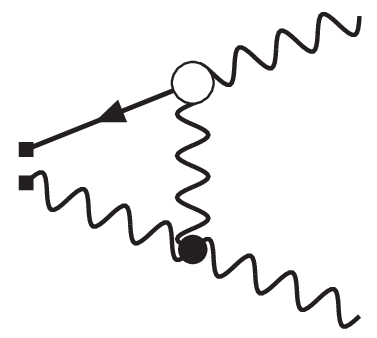}
\caption{}
\label{fig:16}
\end{center}
\end{minipage}
\begin{minipage}{0.3\hsize}
\begin{center}
\includegraphics[width=4cm,clip]{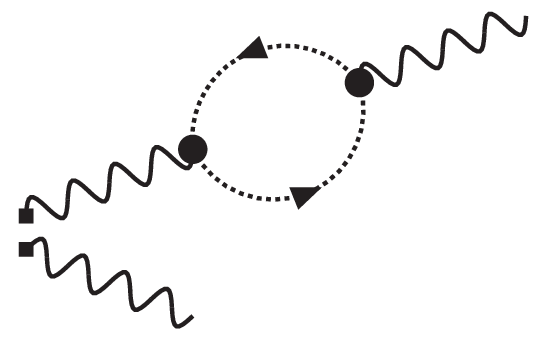}
\caption{}
\label{fig:17}
\end{center}
\end{minipage}
\end{figure}

To find the coefficient functions $F(t)$ and~$G(t)$ in~Eq.~\eqref{eq:(5.1)}, we
consider the correlation function
\begin{equation}
   \left\langle G_{\mu\rho}^a(t,x)G_{\nu\rho}^a(t,x)
   A_\kappa^i(w)A_\omega^j(v)\right\rangle.
\label{eq:(5.8)}
\end{equation}
For~$O(g_0^2)$, there are $17$ flow-line Feynman diagrams
(Figs.~\ref{fig:1}--\ref{fig:17}) that contribute to this correlation function.
In the figures, gauge potentials at the flow time~$t$, $B_\mu(t,x)$, are
represented by small filled squares; the open circle denotes the flow-time
vertex and the full circle is the conventional vertex in the Yang-Mills theory.
We refer the reader to~Ref.~\cite{Luscher:2011bx} for the details of the
Feynman rules for flow-line diagrams.

To read off the coefficient functions $F(t)$ and~$G(t)$ in~Eq.~\eqref{eq:(5.1)}
from the correlation function~\eqref{eq:(5.8)}, we consider the vertex
functions, i.e., amputated diagrams in which the external propagators of the
original Yang--Mills theory are truncated. Therefore, Figs.~\ref{fig:10},
\ref{fig:12} and~\ref{fig:17}, which provide only the conventional wave
function renormalization, should be omitted in the computation of $F(t)$
and~$G(t)$.\footnote{More precisely, these diagrams are different from
conventional Feynman diagrams in that the propagators carry an additional
factor $e^{-tp^2}$ (in the Feynman gauge), where $p$ is the external momentum.
This factor is, however, irrelevant in the present computation of the
coefficients of operators with the lowest number of derivatives.} On the other
hand, the flow-line propagators~\cite{Luscher:2011bx}, the arrowed straight
lines in the diagrams, should not be truncated because these are not
propagators in the quantum field theory but instead represent time evolution
along the flow time.

The tree-level contribution to the vertex function is
\begin{align}
   \text{Fig.~\ref{fig:1}}
   &=\delta_{\rho\sigma}\left[
   \int_{p,q}\,e^{i(p+q)x}\Tilde{A}_\rho^a(p)\Tilde{A}_\sigma^a(q)
   e^{-tp^2}e^{-tq^2}ip_\mu iq_\nu
   \pm(\mu\leftrightarrow\rho,\nu\leftrightarrow\sigma)\right]
\notag\\
   &=F_{\mu\rho}^a(x)F_{\nu\sigma}^a(x)+O(t),
\label{eq:(5.9)}
\end{align}
where
\begin{equation}
   A_\mu(x)=\int_pe^{ipx}\Tilde{A}_\mu(p),\qquad
   \int_p\equiv\int\frac{d^Dp}{(2\pi)^D},
\label{eq:(5.10)}
\end{equation}
and, here and in what follows, the alternating-sign symbol implies
\begin{equation}
   t_{\mu\rho\nu\sigma}\pm(\mu\leftrightarrow\rho,\nu\leftrightarrow\sigma)
   \equiv
   t_{\mu\rho\nu\sigma}-t_{\rho\mu\nu\sigma}-t_{\mu\rho\sigma\nu}
   +t_{\rho\mu\sigma\nu}.
\label{eq:(5.11)}
\end{equation}
This tree-level result was used in obtaining~Eq.~\eqref{eq:(4.20)}.

The vacuum expectation value in the lowest order is
\begin{align}
   \text{Fig.~\ref{fig:2}}&=g_0^2\delta^{aa}\delta_{\rho\sigma}
   \left[\int_\ell\frac{1}{\ell^2}\,e^{-2t\ell^2}
   \ell_\mu\ell_\nu\delta_{\rho\sigma}
   \pm(\mu\leftrightarrow\rho,\nu\leftrightarrow\sigma)\right]
\notag\\
   &=\frac{3}{128\pi^2}(N^2-1)
   g_0^2\frac{1}{t^2}\delta_{\mu\nu}.
\label{eq:(5.12)}
\end{align}

Now, as an example of the computation of one-loop flow-line Feynman diagrams,
we briefly illustrate the computation of~Fig~\ref{fig:13}. A straightforward
application of the Feynman rules in~Ref.~\cite{Luscher:2011bx} in the
``Feynman gauge'' in which the gauge parameters are taken
as~$\lambda_0=\alpha_0=1$, yields the expression,
\begin{align}
   \text{Fig.~\ref{fig:13}}
   &=Ng_0^2\delta_{\rho\sigma}\biggl(
   \int_{p,q}\,e^{i(p+q)x}\Tilde{A}_\alpha^b(p)\Tilde{A}_\beta^c(q)
\notag\\
   &\qquad{}\times
   \int_\ell
   \frac{1}{(p+\ell)^2}\frac{1}{\ell^2}\frac{1}{(q-\ell)^2}\,
   e^{-t(p+\ell)^2}e^{-t(q-\ell)^2}
\notag\\
   &\qquad\qquad{}\times
   i(p+\ell)_\mu i(q-\ell)_\nu
\notag\\
   &\qquad\qquad{}\times
   \left[\delta_{\rho\lambda}(-p-2\ell)_\alpha
   +\delta_{\lambda\alpha}(\ell-p)_\rho
   +\delta_{\alpha\rho}(2p+\ell)_\lambda\right]
\notag\\
   &\qquad\qquad{}\times
   \left[\delta_{\lambda\sigma}(-2\ell+q)_\beta
   +\delta_{\sigma\beta}(-2q+\ell)_\lambda
   +\delta_{\beta\lambda}(q+\ell)_\sigma\right]
\notag\\
   &\qquad\qquad\qquad{}\pm(\mu\leftrightarrow\rho,\nu\leftrightarrow\sigma)
   \biggr).
\label{eq:(5.13)}
\end{align}
To find the coefficients $F(t)$ and~$G(t)$ in~Eq.~\eqref{eq:(5.1)}, we write
this vertex function as
\begin{equation}
   \int_{p,q}\,e^{i(p+q)x}\Tilde{A}_\alpha^a(p)\Tilde{A}_\beta^a(q)
   M_{\mu\nu,\alpha\beta}(p,q),
\label{eq:(5.14)}
\end{equation}
and find the coefficients of
\begin{equation}
   -p_\mu q_\nu\delta_{\alpha\beta}
\label{eq:(5.15)}
\end{equation}
and
\begin{equation}
   -4p\cdot q\delta_{\mu\nu}\delta_{\alpha\beta},
\label{eq:(5.16)}
\end{equation}
respectively, in~$M_{\mu\nu,\alpha\beta}(p,q)$. For this, we first exponentiate
the denominators in~Eq.~\eqref{eq:(5.13)} by using
\begin{equation}
   \frac{1}{(p+\ell)^2}\frac{1}{(q-\ell)^2}
   =\int_0^\infty d\xi\,\int_0^\infty d\eta\,e^{-\xi(p+\ell)^2}e^{-\eta(q-\ell)^2}.
\label{eq:(5.17)}
\end{equation}
We then simply expand the integrand with respect to the external momenta $p$
and~$q$ to $O(p,q)$. The flow-time evolution factor~$e^{-2t\ell^2}$ in the
integrand makes the integral~\eqref{eq:(5.13)} UV finite for any dimension~$D$.
On the other hand, there always exists a complex domain of~$D$ such that the
integral is infrared finite; this provides the analytic continuation of the
integral such that
\begin{align}
   &\int_\ell\frac{1}{\ell^2}\,
   e^{-\alpha\ell^2}=\frac{1}{(4\pi)^{D/2}}\frac{1}{D/2-1}\alpha^{-D/2+1},
\label{eq:(5.18)}
\\
   &\int_\ell\frac{1}{\ell^2}\,e^{-\alpha\ell^2}\ell_\mu\ell_\nu
   =\frac{1}{(4\pi)^{D/2}}\frac{1}{D}\alpha^{-D/2}\delta_{\mu\nu},
\label{eq:(5.19)}
\\
   &\int_\ell\frac{1}{\ell^2}\,e^{-\alpha\ell^2}
   \ell_\mu\ell_\nu\ell_\rho\ell_\sigma
   =\frac{1}{(4\pi)^{D/2}}\frac{1}{2(D+2)}\alpha^{-D/2-1}
   \left(
   \delta_{\mu\nu}\delta_{\rho\sigma}
   +\delta_{\mu\rho}\delta_{\nu\sigma}
   +\delta_{\mu\sigma}\delta_{\nu\rho}
   \right),
\label{eq:(5.20)}
\\
   &\int_\ell\frac{1}{\ell^2}\,e^{-\alpha\ell^2}
   \ell_\mu\ell_\nu\ell_\rho\ell_\sigma\ell_\alpha\ell_\beta
\notag\\
   &=\frac{1}{(4\pi)^{D/2}}\frac{1}{4(D+4)}\alpha^{-D/2-2}
   \left(
   \delta_{\mu\nu}\delta_{\rho\sigma}\delta_{\alpha\beta}
   +\text{$14$ permutations}\right).
\label{eq:(5.21)}
\end{align}
Then it is straightforward to find the coefficients of Eqs.~\eqref{eq:(5.15)}
and~\eqref{eq:(5.16)}, which directly make a contribution to the functions
$F(t)$ and~$G(t)$.

\begin{table}
\caption{The contributions of each flow-line Feynman diagram (in the Feynman
gauge) to the coefficients of Eqs.~\eqref{eq:(5.15)} and~\eqref{eq:(5.16)},
respectively, in the unit of~Eq.~\eqref{eq:(5.22)}. These correspond to the
coefficient functions $F(t)$ and~$G(t)$ in~Eq.~\eqref{eq:(5.1)}.
The numbers with~$*$ are corrected from previous versions of the present paper.
}
\label{table:1}
\begin{center}
\renewcommand{\arraystretch}{2.2}
\setlength{\tabcolsep}{20pt}
\begin{tabular}{crr}
\toprule
 & \multicolumn{1}{c}{$F(t)$}
 & \multicolumn{1}{c}{$G(t)$} \\
\midrule
Fig.~\ref{fig:3}  & $0$ & $0$ \\
Fig.~\ref{fig:4}  & $-3\dfrac{1}{\epsilon}-3\ln(8\pi t)-1$ & $0$ \\
Fig.~\ref{fig:5}  & $-\dfrac{7}{36}$ & $-\dfrac{49}{144}$ \\
Fig.~\ref{fig:6}  & $2\dfrac{1}{\epsilon}+2\ln(8\pi t)-\dfrac{1}{2}$ & $0$ \\
Fig.~\ref{fig:7}  & $\dfrac{19}{288}$ & $\dfrac{121}{384}$ \\
Fig.~\ref{fig:8}  & $\dfrac{35}{96}^*$ & $\dfrac{143}{384}^*$ \\
Fig.~\ref{fig:9}  & $-\dfrac{25}{8}$ & $0$\\
Fig.~\ref{fig:11}  & $\dfrac{1}{3}\dfrac{1}{\epsilon}+\dfrac{1}{3}\ln(8\pi t)-\dfrac{17}{36}$ & $\dfrac{7}{12}\dfrac{1}{\epsilon}+\dfrac{7}{12}\ln(8\pi t)+\dfrac{1}{144}$\\
Fig.~\ref{fig:13}  & $-\dfrac{5}{3}\dfrac{1}{\epsilon}-\dfrac{5}{3}\ln(8\pi t)+\dfrac{25}{36}$ & $-\dfrac{3}{2}\dfrac{1}{\epsilon}-\dfrac{3}{2}\ln(8\pi t)-\dfrac{29}{16}$\\
Fig.~\ref{fig:14} & $3\dfrac{1}{\epsilon}+3\ln(8\pi t)+3$ & $0$ \\
Fig.~\ref{fig:15}  & $3\dfrac{1}{\epsilon}+3\ln(8\pi t)+\dfrac{5}{2}$ & $0$ \\
Fig.~\ref{fig:16}  & $1^*$ & $\dfrac{31}{24}$ \\
$Z$~factors & $-\dfrac{11}{3}\dfrac{1}{\epsilon}$ & $\dfrac{11}{12}\dfrac{1}{\epsilon}$ \\
\bottomrule
\end{tabular}  
\end{center}
\end{table}

In Table~\ref{table:1}, we summarize the contribution of each diagram
computed in the above method in the unit of
\begin{equation}
   \frac{1}{16\pi^2}Ng_0^2.
\label{eq:(5.22)}
\end{equation}
In the last line of the table, ``$Z$~factors'' implies the contributions of the
one-loop operator renormalization factors, $Z_T$~\eqref{eq:(2.13)} and
$Z_M$~\eqref{eq:(2.22)}, through the tree-level diagram, Eq.~\eqref{eq:(5.9)}
(recall Eq.~\eqref{eq:(2.10)}). We see that those operator renormalization
factors precisely cancel the residues of~$1/\epsilon$ and make the coefficients
$F(t)$ and~$G(t)$ finite; this is precisely what we expect from the general
argument. From the results in the table, we then have
\begin{align}
   F(t)&=1+2b_0g^2
   \left[\ln(\sqrt{8t}\mu)+\ln\sqrt{\pi}+\frac{7}{22}\right],
\label{eq:(5.23)}
\\
   G(t)&=-\frac{1}{2}b_0g^2
   \left[\ln(\sqrt{8t}\mu)+\ln\sqrt{\pi}+\frac{1}{11}\right].
\label{eq:(5.24)}
\end{align}
Finally, comparison with the formulas~\eqref{eq:(5.4)}, \eqref{eq:(5.7)},
\eqref{eq:(4.21)} and~\eqref{eq:(4.23)} shows the results quoted
in~Sect.~\ref{sec:4}, Eqs.~\eqref{eq:(4.30)} and~\eqref{eq:(4.31)}. Note that
the coefficients of $\ln(\sqrt{8t}\mu)$ in the explicit one-loop calculation
(Eqs.~\eqref{eq:(5.23)} and~\eqref{eq:(5.24)}) are in agreement with those by
the general argument on the basis of the renormalization group equations and
the trace anomaly (Eqs.~\eqref{eq:(4.21)} and~\eqref{eq:(4.23)}). This
agreement provides a consistency check for our one-loop calculation and
supports the correctness of our reasoning.

\end{document}